\documentclass[openacc]{rsproca_new}

\usepackage{hyperref} 
\usepackage{graphicx}
\usepackage{dcolumn}
\usepackage{bm}

\begin{document}

\title{Droplet impact on immiscible liquid pool: Multi-scale dynamics of entrapped air cushion at short timescales}

\author{
Durbar Roy$^{1}$, Sophia M$^{1}$, Srinivas Rao S$^{1}$ and Saptarshi Basu$^{1}$}

\address{$^{1}$Department of Mechanical Engineering, Indian Institute of Science, Bengaluru, KA 560012, India}

\subject{Mechanical Engineering, Fluid Mechanics}


\corres{Saptarshi Basu\\
\email{sbasu@iisc.ac.in}}

\begin{abstract}
We have detected unique hydrodynamic topology in thin air film surrounding the central air dimple formed during drop impact on an immiscible liquid pool. The pattern resembles spinodal and finger-like structures typically found in various thin condensed matter systems. However, similar structures in thin entrapped gas films during drop impacts on solids or liquids have not been reported to date. The thickness profile and the associated dewetting dynamics in the entrapped air layer are investigated experimentally and theoretically using high-speed reflection interferometric imaging and linear stability analysis. We attribute the formation of multiscale thickness perturbations, associated ruptures, and finger-like protrusions in the draining air film as a combined artifact of thin-film and Saffman-Taylor instabilities. The characteristic length scales depend on the air layer dimensions, the ratio of the liquid pool to droplet viscosity, and the air-water to air-oil surface tension.
\end{abstract}
\maketitle

\section{Introduction}
Drop impacts on liquid pools are ubiquitous in many industrial, biological, agricultural, and medical applications [1-3].
The droplet and the liquid pool deforms prior to impact [4-6] due to the lubrication pressure of the entrapped air between the pool and the droplet (Fig. 1(a)).  
Entrapped air layer dynamics beneath the droplet provide insights into various types of bubble formation mechanisms like Messler entrainment, bubble chandeliers, and hanging necklaces [7,8] (Fig. 1(b)). Various research groups [9-11] pursue drop impact physics experimentally and theoretically due to its application in 3D printing, spray painting, and novel cooling technologies [11-13]. Air bubbles of various sizes that get entrapped during spray coating are detrimental to superior surface finish. On the other hand, air bubbles trapped by raindrops provide a mechanism of gaseous exchange between the atmosphere and natural water bodies that support a wide variety of aquatic life forms [14].

Bubble entrapment mechanisms during various kinds of drop impacts on solids and liquids were studied by many groups [14]. The bubble entrapment mechanism for drop impact on pools is related to the \emph{entrapped air layer dynamics} that forms the central theme of our work.
The outcome of drop impact dynamics on solids and liquids depends on numerous parameters like fluid properties, impact velocity, atmospheric pressure, plate temperature, and droplet size, to name a few.
The pool depth is also an essential parameter for impact on liquid pools. Hicks et al. [4,5] studied the effect of liquid pool height on air cushioning and splash dynamics for drop impact. Hicks and co-workers showed that a critical balance occurs when the pool depth is of the same order of magnitude as the horizontal extent of interactions between the pool and drop. The air layer dynamics do not depend on the pool height and pool base for significant depth (deep pool limit). On the contrary, air layer dynamics during drop impact on a solid surface can be thought of as a limiting case of drop impact on liquid pools with negligible depth compared to drop radius (shallow pool limit).  The entrapped air layer dynamics lie between deep and shallow pool limits for all impact conditions. The rise in pressure deforms the droplet forming the \emph{central dimple} prior to impact (Fig. 1(a)). Lubrication pressure of the entrapped air is inversely proportional to the air layer thickness and hence increases as the droplet approaches the liquid pool. The effect of lubrication pressure decreases in the radial direction forming a thin air layer region surrounding the central air dimple, which we refer to as the \emph{peripheral air disc} (Fig. 1(a)). The air thickness near the dimple is relatively higher ($\mathcal{O}(10^1)$) than the peripheral air disc region. In general, the air disc curves radially upwards for impact on liquids. However, the curvature depends on the impact Weber number ($We={\rho}_wV_0^2R_0/{\sigma}$) and liquid pool properties like viscosity, where ${\rho}_w$ is the density of the impinging droplet (water here), $V_0$ is the impact velocity, $R_0$ is the radius of the impinging droplet and ${\sigma}$ is liquid-air surface tension. High impact Weber number and low liquid pool viscosity result in significant air disc curvature [15]. 
However, impact on highly viscous liquid pool like silicone oil at relatively small Weber number ($We<15$), the curvature reduces significantly. Therefore, the peripheral air disc can be approximated as a flat region (a tangent plane approximation). 
Researchers have studied the central dimple extensively [15-17] in the context of impacts on liquids and solids. However, the nature of the peripheral air disc and its associated dewetting dynamics is relatively unknown.

\begin{figure*}
  \centerline{\includegraphics[scale=1]{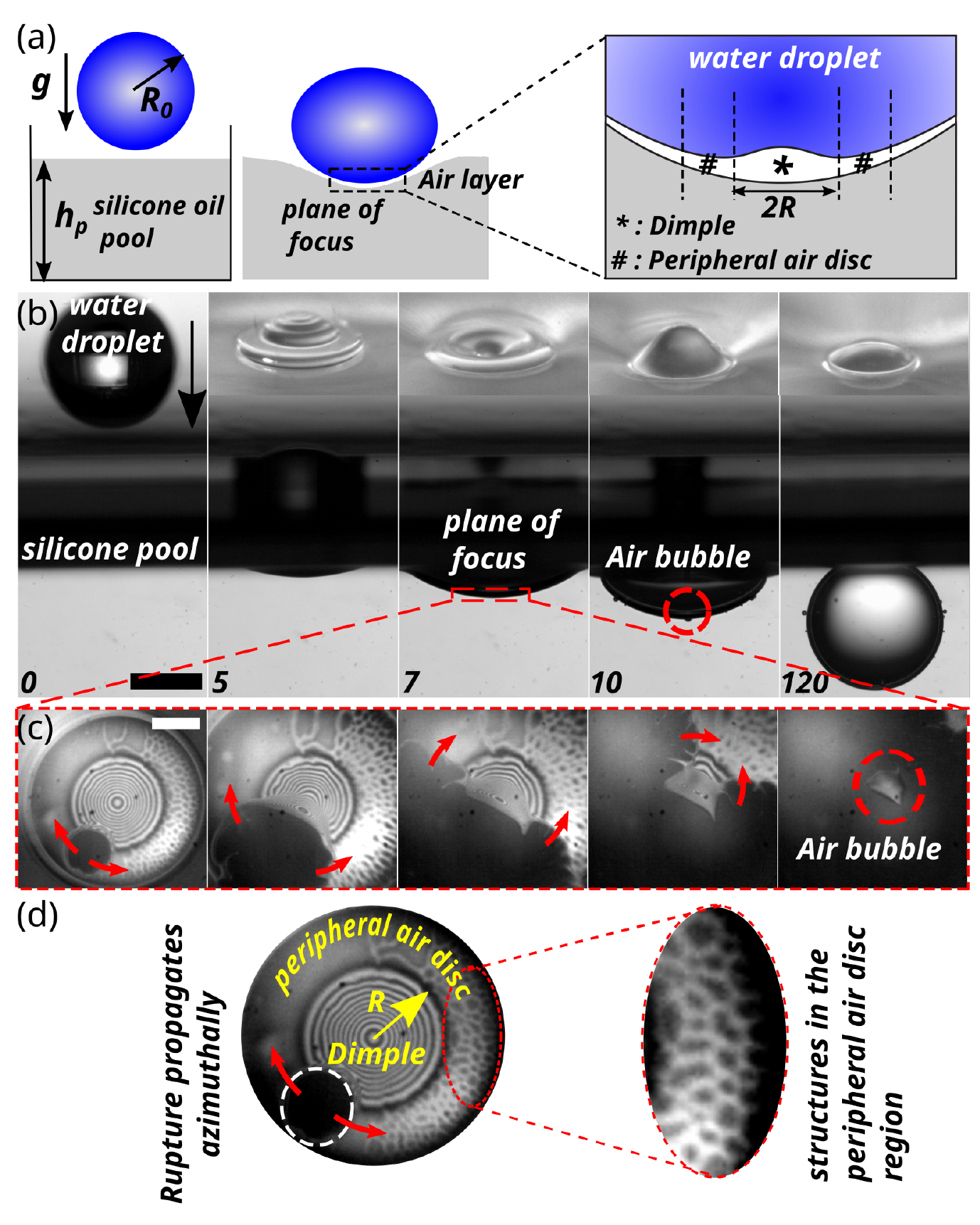}}
  \caption{Overview of the various physical processes beneath an impacting water droplet on silicone oil.
    (a) Schematic representation of the air layer between the impinging water droplet (blue color) and the silicone oil pool (grey color). $g$ represents acceleration due to gravity, and $R_0$ represents the impinging droplet radius. The plane of focus marked by the dotted rectangle is shown as an expanded view on the right. Dimple and peripheral air disc are represented as * and \#, respectively. The dimple diameter is represented as $2R$.
    (b) Time series snapshots during the impact on silicone pool. The timestamps are in milliseconds. The black scale bar represents 1mm. Air bubbles (red dotted circle) can be observed at 10 milliseconds and beyond.
    (c) High-speed reflection interferometric time-series images of the air layer beneath the impinging droplet. The air layer rupture propagates in the azimuthal direction (red arrows) and forms the air bubble (red dotted circle).
    The white scale bar represents 250 microns.
    (d) Magnified view of the air entrapped between the water droplet and the silicone pool. The air layer consists of two regions, the dimple of radius R and the peripheral air disc.}
\label{Figure1}
\end{figure*}

In this study, we investigate the air layer dynamics in the central dimple and peripheral disc region for relatively low impact Weber number ($We{\sim}\mathcal{O}(10)$) on silicone oil pool of three different viscosities using experimental and theoretical scaling analysis.
The entrapped air forms two distinct interfaces for drop impacts on immiscible liquids, i.e., top interface with the impacting droplet (water) and bottom interface with liquid pool (silicone oil). The draining of the entrapped air due to the squeezing action of the impinging droplet can be formulated as a dewetting process.
As the entrapped air layer thickness ($h$) in the disc region reduces to $h{\sim}100nm$, the air layer dewetting/draining dynamics are affected by molecular interaction potentials like van der Waals attraction and double layer repulsion [18]. New fundamental length scales like spinodal structures [19] can exist in meta-stable state due to the balance of capillary and intermolecular forces. Such spinodal dewetting patterns were observed in condensed matter systems like thin polymers, liquid crystals, and liquid metal films [20-24].
However, dewetting patterns in gaseous media (air here) in drop impact studies are relatively unknown in the scientific community. We propose the topology observed in the peripheral air disc results as a combined effect of thin-film/spinodal instability and Saffman Taylor instability.
Saffman Taylor/viscous fingering are formed at an unstable interface of differential viscosity where the less viscous fluid pushes a high viscous fluid.
Viscous fingering have been intensively studied in the fluid dynamics community [25-29]. The mechanism of viscous fingering is dominantly observed in porous media and drainage processes. Finger-like structures near the vicinity of the central air dimple during drop impacts have not been reported to date.
We have observed hydrodynamic structures in the dewetting peripheral air disc surrounding the central dimple that resembles the finger-like structures.
We attribute the instability structures as a combined artifact of thin-film [18] and Saffman-Taylor instabilities [25]. The peripheral air disc thins out due to pressure-induced height fluctuations similar to the dewetting mechanism in thin polymer films [20].
This is because thin-film instability causes the interface to develop perturbations, which further act as a scaffolding for Saffman-Taylor instability to form finger-like structures that propagate in the radial direction.

\begin{figure}
  \centerline{\includegraphics{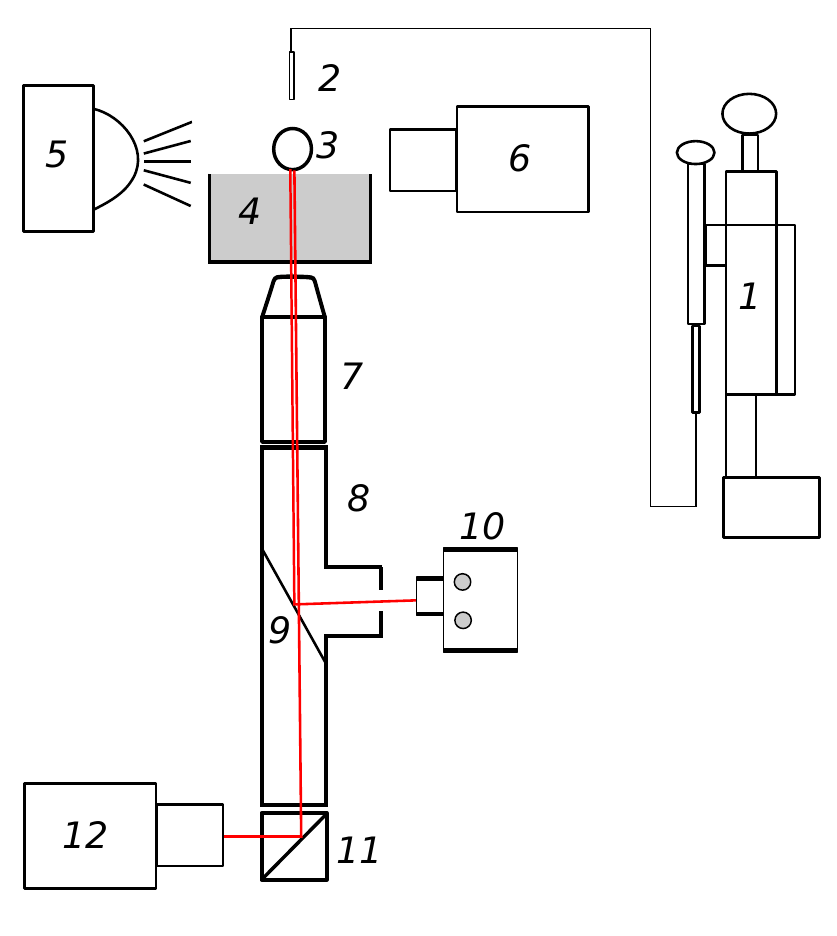}}
  \caption{Schematic of the experimental setup. Various components of the setup are labelled by numeric values. 1:Syringe pump, 2: Needle, 3: Water droplet, 4: Silicone oil liquid pool, 5: Backlight source for shadowgraphy, 6: High speed camera, 7: Microscope objective, 8: Zoom lens, 9: Beam splitter, 10: Laser source, 11: Dichoric mirror, 12: High speed camera}
\label{Figure2}
\end{figure}

Fig. 1(a) shows a schematic representation of the air layer entrapped between an impinging water droplet and silicone oil pool. The pressure in the entrapped air layer increases as the droplet approaches the liquid pool. The increased pressure in the air layer results in a dimple formation. The entrapped air layer is divided into two connected regions, the central dimple (marked as *) and the peripheral air disc (marked as \#). New topological structures were detected in the peripheral air disc region.
Fig. 1(b) depicts the impact event captured using high-speed shadowgraphy and perspective view imaging. The plane of focus of the reflection interferometry setup is marked with a dotted red rectangle. Fig. 1(c) shows high-speed interferogram snapshots that reveal the hydrodynamic structures in the peripheral air disc. The azimuthal air layer ruptures and propagates (marked by a red arrow), forming the central air bubble (marked by a red dotted circle).
Fig. 1(d) shows the magnified view of the air disc where the dimple radius is represented as $R$. The lack of left-right symmetry in the air disc is due to slight asymmetry in air layer entrapment during drop impact. Refer to \href{https://drive.google.com/file/d/1p_a8bppg7aVLDHQDb3JuG-JhQQoOXPDZ/view?usp=sharing}{Fig. S1} in the supplementary text for an example of symmetric patterns observed throughout the peripheral air disc.

\section{Materials and Methods}
 \subsection{Experimental setup}
 Figure 2 shows the experimental setup. It consists of two high-speed cameras (Photron SA5), a pulse-diode laser source for interferometric illumination (Cavitar Cavilux smart UHS,
400 W power, ${\lambda}=632nm$), a white light source for shadowgraphy, interferometric optics (including a Navitar zoom lens, dichoric mirror, and beam splitter), a microscope objective (5x, 10x, and 20x), a liquid pool of silicone oil and a syringe pump (New Era Pump Systems, NE-1010) for generating droplet of radius $R_0$ ${\sim}1mm$.
We used high-speed reflection interferometric imaging at 10000 - 50000 frames per second using a monochromatic diode laser source (632 nm). A spatial resolution of 1.5 micrometers per pixel was used to resolve the constructive and destructive interference patterns. We have explored the air layer dynamics at the low impact Weber number regime ($We{\sim}10$) due to the dominant effect of the air cushion. The droplet was allowed to fall freely under gravity to achieve a particular impact Weber number.
In this study, we focus on the effect of liquid pool viscosity on the air layer dynamics at a constant impact Weber number of approximately ($We{\sim}10$).
Deionised water droplet of radius $1.1mm$ was used as the impinging droplet, while silicone oil of three different kinematic viscosities of ($5$, $350$, and $950cSt$) was used for the pool. The silicone pool height (refer to Fig. 1(a)) was maintained at approximately five times the droplet radius ($h_p{\sim}5R_0$). We used high-speed interferograms to evaluate the air layer thickness profile.

\subsection{Evaluation of air layer thickness profile from interferograms}
A monochromatic light beam from the diode source propagates through the microscope objective(20X) and travels towards the silicone oil free surface. A part of the incident light gets reflected at the air-oil interface to register an intensity $I_1$; the transmitted light travels through the air layer between the droplet and pool, causing another reflection from the bottom surface of the impinging droplet that gives rise to intensity of $I_2$. Lights of intensity $I_1$ and $I_2$ (with a phase shift due to the additional optical path travelled) interferes and forms a resultant intensity pattern $I$  expressed in the form [30-32]
\begin{equation}
    I(x_1,x_2)=I_1+I_2+2\sqrt{I_1I_2}cos({\phi}(x_1,x_2))
\end{equation}
where, ${\phi}(x_1,x_2)$ is the phase shift.
The interference patterns are captured using high-speed cameras. Sample interferograms are shown in Fig. 3.
Here $x_1,x_2$ represent the acquired two-dimensional image coordinates. 
The main feature of equation (2.1) is to quantify the fringe patterns depending on the variation of a cosine-modulated function. Hence, if the phase gradient is high, high-density fringes are observed. Similarly, a smaller phase gradient corresponds to low fringe density. In this context, the developed fringe patterns during droplet impingement have been processed by employing the fast-frequency guided sequential demodulation (FFSD) method [33,34]. Fast-frequency guided algorithms are suitable for transient phenomena and consequent comprehensive extraction of the two-dimensional phase field of the complex fringe patterns. 
The gradient of phase distribution for the obtained fringe patterns has been estimated from the fast-frequency guided sequential demodulation given as [33]:
\begin{equation}
    {\phi}(x_1,x_2)={\omega}(x_1,x_2)e^{i{\theta}(x_1,x_2)}
\end{equation}
Here, ${\omega}(x_1,x_2)$ signifies the amplitude to determine the local frequencies precisely in the spatial domain through frequency-guided strategy, and  ${\theta}$ denotes the angle of the complex number determined from the initial intensity variables. The obtained phase variables are constructed to be continuous and unwrapped. The height of the air-layer profiles can be estimated for the preferred wavelength of the light source ${\lambda}$ and refractive index $n$ of the fluid medium as [35]:
\begin{equation}
    h(x_1,x_2)=\frac{\lambda}{2{\pi}n}{\phi}(x_1,x_2)
\end{equation}

\section{Results}
The air layer beneath the impacting droplet on a liquid pool consists of two distinct connected regions, a central dimple, and the peripheral air disc (Fig. 3(a) and 3(b)).
The dimple radius $R$, height $h_0$ and the initial radial air velocity was observed to be the largest for impact on $5cSt$ followed by $350$ and $950cSt$. However the peripheral air disc was the smallest for $5cSt$, followed by $350$ and $950cSt$ (Fig. 3(c), 3(d) and 3(e) shows the interferogram snapshots for impact on $5$, $350$ and $950cSt$ silicone oil).
The peripheral air disc radius $r_0(t)$ dependence on time was found to be $t^{2/3}$ as a result of small Capillary number ($Ca={\mu}_aV_0/{\sigma}_{aw}$) where ${\mu}_a$ is the air viscosity, $V_0$ is the impact velocity and ${\sigma}_{aw}$ is the air-water surface tension. Further the peripheral air disc thickness was one order smaller than the central air dimple thickness. The air disc on $5cSt$ was highly unstable compared to $350$ and $950cSt$ resulting in very fast ruptures (compare the time scales of Fig. 3(c), 3(d) and 3(e)). 
Fig. 4(a) shows the region of focus for capturing the interference signals. Fig. 4(b) shows the intensity map of the interference signals plotted as a 3D surface. Unique spinodal and finger-like  structures in the peripheral air disc was observed for drop impact on $350$ and $950cSt$ at impact Weber number of $We{\sim}\mathcal{O}(10)$ using high-speed reflection interferometric imaging (Fig. 4(b), 4(c) and 4(d)). We propose that the structures in the peripheral air disc result due to the combined effect of thin-film/spinodal and Saffman Taylor instabilities. Scaling and linear stability analysis was used to unearth the air layer dynamics and the related instabilities beneath the impacting droplet. The length scales of experimentally measured structures (Fig. 4(d)) conforms to the most unstable wavelength of thin film (${\lambda}_{TF}{\sim}\mathcal{O}(11.26{\mu}m)$) and Saffman Taylor instability (${\lambda}_{ST}{\sim}\mathcal{O}(13.89{\mu}m)$). (Refer to the supplementary videos \href{https://drive.google.com/file/d/1p_a8bppg7aVLDHQDb3JuG-JhQQoOXPDZ/view?usp=sharing}{SV1, SV2 and SV3} links in the supplementary article provided for the high speed interferograms visualizing the entrapped air layer for $5$, $350$ and $950cSt$ respectively.)

\begin{figure*}
  \centerline{\includegraphics{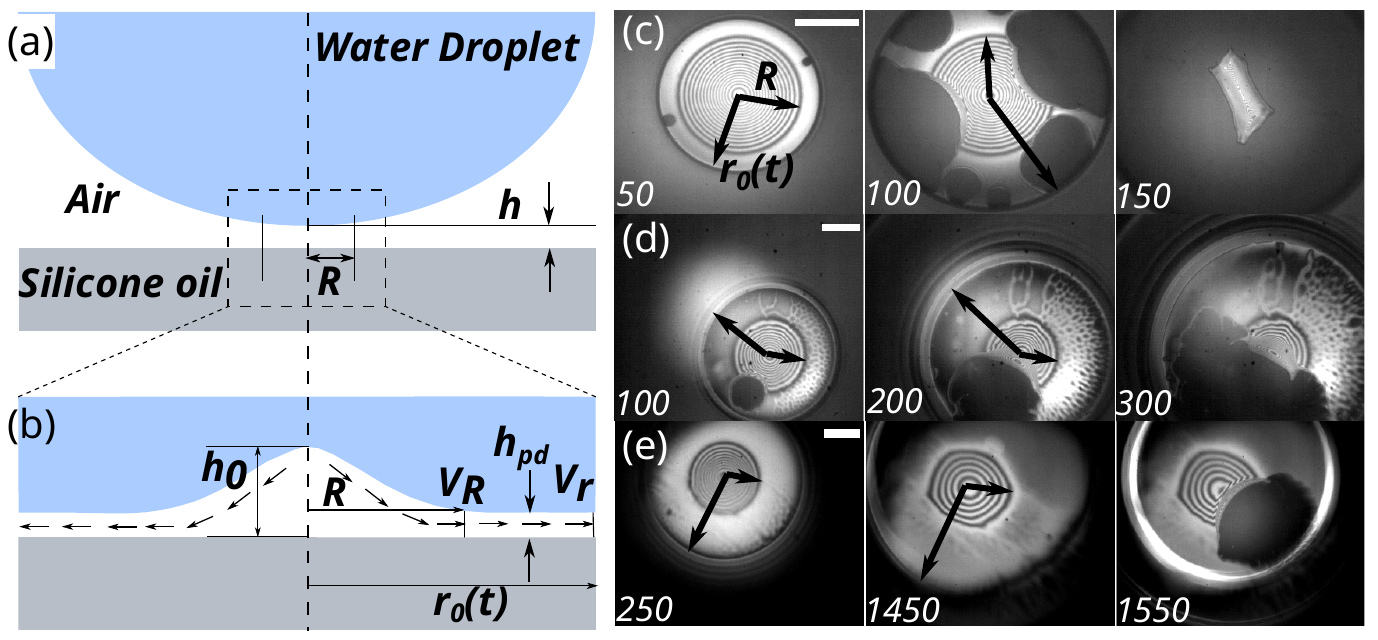}}
  \caption{ (a) Schematic representation of the droplet impact phenomena on the liquid pool just prior to the formation of the central dimple.
  (b) Magnified view of the fully formed central dimple and the peripheral air disc. The relevant dimensions are annotated. $h_0$ denotes the dimple height, $R$ represents the dimple radius, $h_{pd}$ represents the thickness of the peripheral air disc.
  High-speed reflection interferometric image sequence depicting peripheral air disc expansion.
The scale bar in white represents $250{\mu}m$. The relative timestamps are in microseconds. The dimple radius is represented as $R$; expanding peripheral air disc radius is represented as $r_0(t)$. 
    (c) Impact on 5 cSt silicone oil pool. 
    (d) Impact on 350 cSt silicone oil pool.
    (e) Impact on 950 cSt silicone oil pool.}
\label{Figure3}
\end{figure*}

\section{Discussions}
The droplet on approach towards the liquid pool displaces the air entrapped in a thin region (Fig. 1(a)). Two distinct connected regions exist and characterize the air layer, the central dimple and the peripheral air disc surrounding the dimple. The air layer dynamics in both regions are presented in the following subjections (subsection (a) and (b)) using scaling analysis .

\subsection{Dynamics of the entrapped air layer (Dimple region)}
Fig. 3(a) schematically represents the air layer as the droplet approaches the liquid pool. The impacting water droplet is shown in blue, the silicone oil pool in gray, and the entrapped air layer in white. 
A cylindrical control volume beneath the impacting droplet of radius $R$ and height $h$ was chosen for the subsequent analysis. 
Applying conservation of mass in the control volume assuming air to be in-compressible (low Mach number) and having a constant density, we have 
\begin{equation}
    V_0{\pi}R^2 = V_R2{\pi}Rh
\end{equation}
where $V_0$ is the impact velocity, $R$ is a radial length scale characterizing the air layer and represents the dimple radius, $V_R$ is the air velocity at the edge of the dimple at the initial phase of expansion.
The radial air velocity $V_R$ at a radial distance $R$ from the centre could be expressed in terms of the impact velocity $V_0$, length scales $h$ and $R$ as
\begin{equation}
    V_R = \frac{V_0R}{2{h}}
\end{equation}
Note however that the scale of $h$ is still unknown.
As the droplet approaches, the pressure in the air layer increases resulting in the formation of the dimple as shown in Fig. 3(b).
The thin entrapped air region can be modelled using the Stokes equation to first order accuracy as was shown by Hicks et al. [4]
\begin{equation}
   {\mu}_a{\nabla}^2\bm{V}{\sim}\bm{\nabla}p_a
\end{equation}
As the air layer thickness is relatively small compared to the droplet radius, lubrication approximation (ref) is valid for the entrapped air layer.
The dominant forces in the entrapped air layer region are the viscous forces and the pressure gradient.
The radial component of equation (4.3) under the lubrication approximation can be written as
\begin{equation}
    \frac{{\partial}^2V_r}{{\partial}z^2}{\sim}\frac{1}{{\mu}_a}\frac{{\partial}p_a}{{\partial}r}
\end{equation}
where the $V_r$ denotes the radial air velocity, $z$ represents the axial coordinate orthogonal to the radial direction, $p_a$ represents the air pressure in the entrapped air layer, ${\mu}_a$ represents the air viscosity and $r$ denotes the radial coordinate. 
Rewriting equation (4.4) in terms of the dominant scales ($V_r{\sim}V_R$, $z{\sim}h$, $p_a{\sim}{\Delta}p_a$) we have
\begin{equation}
    \frac{{\Delta}p_a}{R}{\sim}\frac{{\mu}_aV_R}{h^2}
\end{equation}

Using the scale of $V_R$ from equation (4.2) in equation (4.5) and solving for the scale of ${\Delta}p_a$ we have

\begin{equation}
    {\Delta}p_a{\sim}\frac{{\mu}_aV_0R^2}{2h^3}
\end{equation}
Note that the pressure in the thin air film region scales as ${\Delta}p_a{\sim}h^{-3}$. Therefore, as the air film thickness decreases, the pressure increases. For a certain critical height $h=h_*$ the air pressure becomes comparable to the capillary pressure ($2{\sigma}_{aw}/{R_0}$) at the air water interface that leads to the formation of an air dimple. Therefore at the critical air layer thickness the equivalence of pressure  scales leads to 
\begin{equation}
    \frac{{\mu}_aV_0R^2}{2h^3_{*}}{\sim}\frac{2{\sigma}_{aw}}{R_0}
\end{equation}
From equation (4.7) the scale of $h_{*}$ can be evaluated as
\begin{equation}
    h_{*}{\sim}{\left[\frac{{\mu}_aV_0R^2R_0}{4{\sigma}_{aw}}\right]^{1/3}}
\end{equation}
where ${\sigma}_{aw}$ is the air water surface tension. Using the definition of capillary number $Ca={\mu}_aV_0/{\sigma}_{aw}$, equation (4.8) can be written as 
\begin{equation}
    h_*{\sim}{\left[\frac{CaR^2R_0}{4}\right]^{1/3}}
\end{equation}
 
Equation (4.9) implies the existence of a critical air layer thickness at which the pressure rise in the air layer surpasses the capillary pressure of the approaching water droplet causing a dimple to form in the centre of the droplet. Using the numerical values of impact velocity $V_0{\sim}\mathcal{O}(0.5m/s)$, air viscosity ${\mu}_a{\sim}\mathcal{O}(1.81{\times}10^{-5}kg/ms)$, air-water surface tension ${\sigma}_{aw}{\sim}\mathcal{O}(0.072N/m)$, dimple radius $R{\sim}\mathcal{O}(200{\times}10^{-6}m)$ and droplet radius of $R_0{\sim}\mathcal{O}(1.1{\times}10^{-3}m)$, the critical thickness is $h_{*}{\sim}\mathcal{O}(2.17{\times}10^{-6}m)$.
The Capillary number for $V_0{\sim}\mathcal{O}(0.5m/s)$ is $Ca{\sim}\mathcal{O}(1.26{\times}10^{-4})$ signifying surface tension effects are dominant compared to viscous effects in the entrapped air layer.
Note that the dimple height at the centre $h_0$ will be of the same order of magnitude as $h_{*}$, however $h_0{\gtrsim}h_{*}$ due to formation of the central air dimple.
Using the value of $h_{*}$ from equation (4.8) in equation (4.2) the initial radial air velocity scale at the edge of the central dimple ($r=R$) becomes
\begin{equation}
    V_R{\sim}\frac{V_0R}{2}{\left[\frac{{\mu}_aV_0R^2R_0}{4{\sigma}_{aw}}\right]^{-1/3}}
\end{equation}
Rearranging the above equations and non-dimensionalizing with respect to the impact velocity $V_0$ we have
\begin{equation}
    \frac{V_R}{V_0}{\sim}{\left[\frac{1}{2Ca}\frac{R}{R_0}\right]^{1/3}}
\end{equation}
 The initial radial air velocity at the edge of central dimple is $V_R{\sim}\mathcal{O}(4.77m/s)$ and the order of magnitude conforms with the experimental peripheral air disc velocity. The experimentally measured initial peripheral air disc velocity for $5{\:}cSt$, $350{\:}cSt$ and $950{\:}cSt$ is of the order of $5.32m/s$, $2.52m/s$ and $1.71m/s$ respectively.
 
 \begin{figure}
  \centerline{\includegraphics{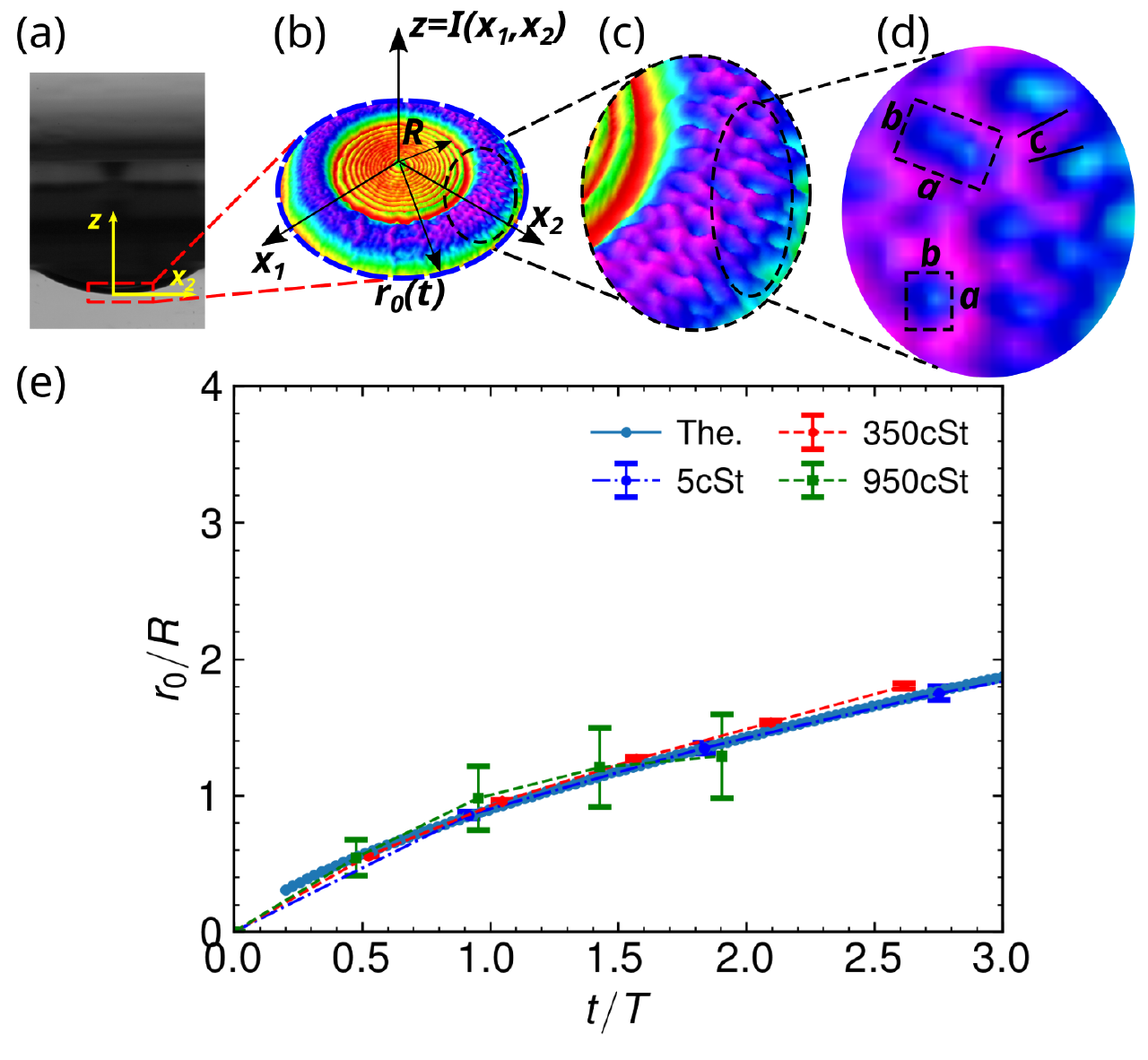}}
  \caption{(a) Region of focus for bottom view interferometric imaging shown as a dotted red rectangle.
    (b) 3D visualization of the reflection interferometric signal (False color)
    (c) Zoomed in 3D view of the peripheral air disc region
    (d) Relevant experimental length scales in the peripheral air disc region. $a,b$ experimentally characterizes the length scales over which the air thickness height changes. The finger like structures observed at the advancing front of the peripheral air disc can be characterized experimentally using the finger width $c$.
    (e) Comparison of the experimental and theoretical air disc expansion characteristics (peripheral air disc radius $r_0$) for drop impact on $5$, $350$ and $950cSt$ silicone oil. $R$ denotes the dimple radius and $T$ denotes the time interval during which the dimple radius $R$ becomes approximately constant.}
\label{Figure4}
\end{figure}

\subsection{Dynamics of the entrapped air layer (Expansion of the Peripheral air disc)}
As the peripheral air disc radius becomes larger than the dimple radius ($r(t)>R$), the air disc expands due to the effect of surface tension forces at the top (air-water) and bottom (air-silicone) interface.
This is due to a very low value of Capillary number ($Ca{\sim}\mathcal{O}(1.26{\times}10^{-4})$) signifying surface tension effects are stronger than viscous effects. Newton second law applied to the expanding air disc gives
\begin{equation}
    {\rho}_a\frac{d^2r_0}{dt^2}{\sim}\frac{2({\sigma}_{aw}+{\sigma}_{as})}{L}\frac{d^2h}{dr^2}
\end{equation}
where ${\rho}_a$ is the air density, ${\sigma}_{aw}$ is the air-water surface tension, ${\sigma}_{as}$ is the air silicone oil surface tension, $h$ represents the air layer height profile as a function of radial coordinate $r$ and $L$ is a characteristic length scale orthogonal to the radial coordinate and related to the air layer geometry. The left hand side of equation (4.12) represents the inertial terms per unit volume and right hand side represents the surface tension forces along the air-water and air-silicone interface. Rewriting the scaling form of equation (4.12) using the dominant scales ($r{\sim}r_0$, $L{\sim}h{\sim}h_{pd}$) where $h_{pd}$ is the thickness of the peripheral air disc
\begin{equation}
    \frac{{\rho}_ar_0}{t^2}{\sim}\frac{2({\sigma}_{aw}+{\sigma}_{as})h_{pd}}{h_{pd}r_0^2}
\end{equation}
Rearranging equation (4.13) and solving for the scale of $r_0$ we have
\begin{equation}
    r_0{\sim}{\left(\frac{2({\sigma}_{aw}+{\sigma}_{as})}{{\rho}_a}\right)^{1/3}}t^{2/3}
\end{equation}
Non-dimensionalizing equation (4.14) with respect to the dimple radius $R$ gives us
\begin{equation}
    \frac{r_0}{R}{\sim}{\left(\frac{2({\sigma}_{aw}+{\sigma}_{as})T^2}{{\rho}_a{R^3}}\right)^{1/3}}{\left(\frac{t}{T}\right)^{2/3}}={\Gamma}{\left(\frac{t}{T}\right)^{2/3}}
\end{equation}
where $R$ is the dimple radius, $T$ is the time scale at which the peripheral air disc radius becomes constant and 
\begin{equation}
    {\Gamma}={\left(\frac{2({\sigma}_{aw}+{\sigma}_{as})T^2}{{\rho}_a{R^3}}\right)^{1/3}}   
\end{equation}
Fig. 4(e) shows the comparison of the experimental peripheral air disc radius evolution with the theoretical scaling predicted by equation (4.15). The experimental value of ${\Gamma}{\sim}\mathcal{O}(0.9)$ where as the theoretical scale of ${\Gamma}$ for $5{\:}cSt$, $350{\:}cSt$ and $950{\:}cSt$ are $2.81$, $4.95$ and $7.44$.
It can be inferred from Fig. 4(e) that the experimental scale conforms with the theoretical scale
predicted by equation (4.15) and by the fact that the numerical value of ${\Gamma}{\sim}\mathcal{O}(1)$ experimentally as well as theoretically.

\begin{figure}
  \centerline{\includegraphics{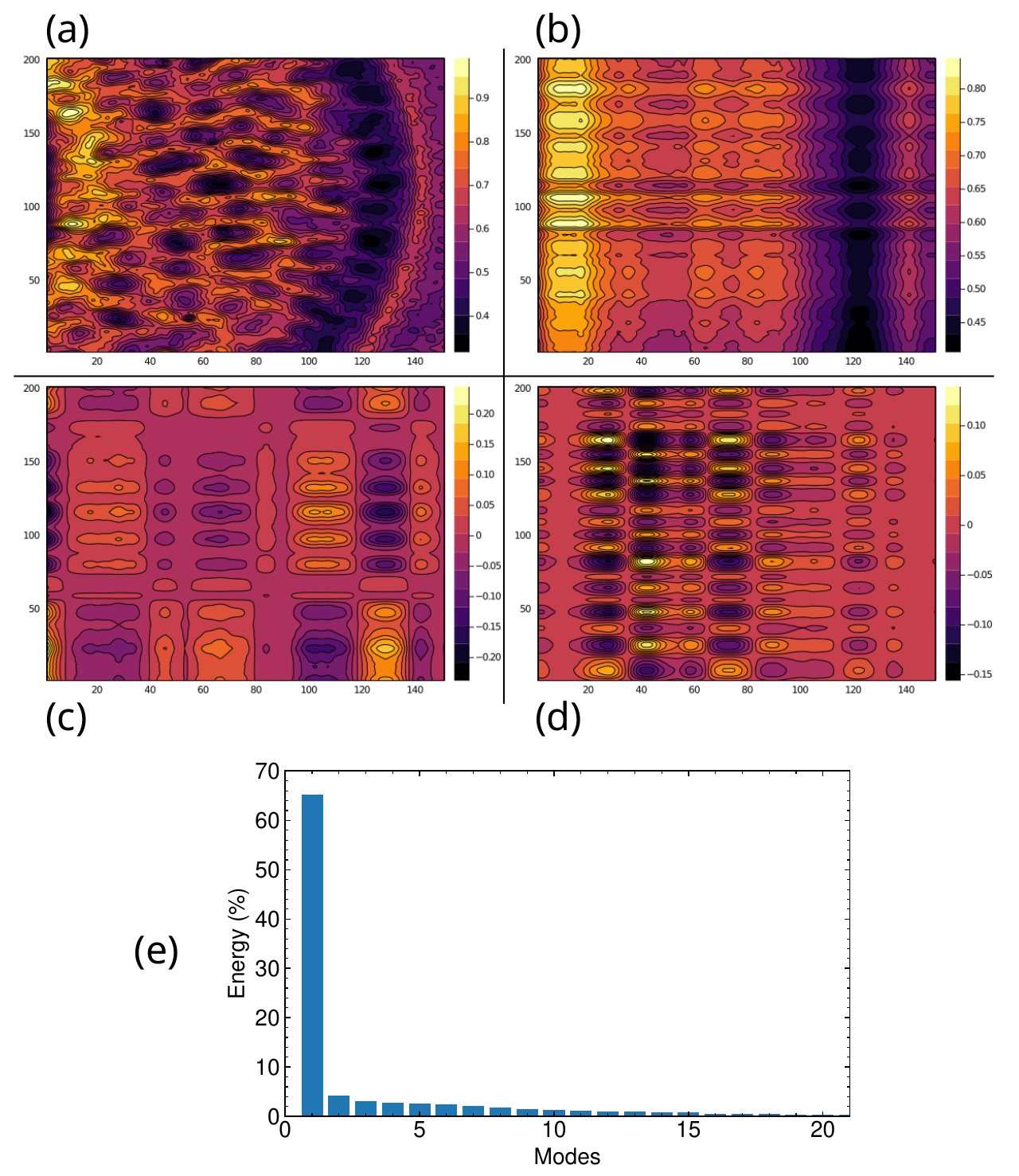}}
  \caption{SVD based Proper orthogonal decomposition (POD) modes [36] of the peripheral air disc region computed using singular value decomposition method. (a) Input image used for extracting the proper orthogonal decomposition modes. (b) POD mode 1
   (c) POD mode 2 (d) POD mode 3. (e) Percentage energy of various POD modes evaluated based on the singular values of the input image.}
\label{Figure5}
\end{figure}

\subsection{Structures in the peripheral air disc region}
\subsubsection{Physical mechanism for structure formation in the peripheral air disc}
The peripheral air disc expands in the radial direction according to equation (4.15).
As was discussed in the previous sections, unique hydrodynamic structures were observed in the peripheral air disc (Fig. 4(d)). In order to characterize the structures present in the peripheral air disc, the spatial intensity distribution of the interferograms was decomposed into orthogonal spatial modes using SVD (singular value decomposition) based proper orthogonal decomposition method [36,37]. The POD was performed to identify and distinguish various length scales present in the disc region. We found that the structures in the peripheral air disc are layered in a very thin region of the order of the dimension of the fingers. Fig. 5(a) shows the input image to the POD algorithm. The input image corresponds to the peripheral air disc region surrounding the central air dimple. The first three dominant modes are shown in Fig. 5(b), (c), and (d), respectively. The POD modes are rank-ordered, with the first mode being the most dominant. Fig. 5(e) shows the percentage energy distribution in various modes. The energies of individual modes were evaluated from the singular values. The first three modes mimic the essential length scales and their corresponding distributions (approximately 75\% of the total energy).
As it can be inferred from the various decomposed modes, the peripheral air disc has intensity fluctuations that have both radial and azimuthal components. As the air layer drains out, the thickness in the peripheral air disc reduces to less than one micron. At these small scales, peripheral air disc thickness $h_{pd}$ exists as a balance between the pressure in the air layer with the disjoining pressure. The disjoining pressure scales as $A/h^3$ where $A$ is the Hamackers constant. The pressure in the air layer scales as ${\sigma}_{aw}/R_0$ as discussed in the previous section.
Therefore the balance of disjoining and the air pressure leads to
\begin{equation}
    \frac{A}{h_{pd}^3}{\sim}\frac{{\sigma}_{aw}}{R_0}
\end{equation}
Solving for the scale of $h_{pd}$ we have
\begin{equation}
    h_{pd}{\sim}{\left[\frac{AR_0}{{\sigma}_{aw}}\right]}^{1/3}
\end{equation}
Using the value of $A{\sim}\mathcal{O}(10^{-19})$ [38], the peripheral air disc thickness scale predicted from equation (4.18) is $h_{pd}{\sim}\mathcal{O}(1.1{\times}10^{-7}m)$ which agrees on an order of magnitude level with the experimental measured value of $\mathcal{O}(3.3{\times}10^{-7}m)$, $\mathcal{O}(3.8{\times}10^{-7}m)$ and $\mathcal{O}(4.1{\times}10^{-7}m)$ for impact on $950cSt$, $350cSt$ and $5cSt$ respectively.

Fig. 6 depicts the schematic representation showing the structure formation mechanism in the peripheral air disc. The peripheral air disc thickness is smaller than the central dimple thickness by order of magnitude, i.e., $h_{pd}/h_0{\sim}\mathcal{O}(10^{-1})$ (by comparing the order of magnitude of equation (4.18) and equation(4.9)). As the entrapped air drains out, small perturbations develop and trigger an instability (Fig. 6(a)). The waves are present in both the top as well as the bottom interface of the air layer; however, the air-silicone interface becomes more unstable due to the lower value of surface tension compared to air-water interface. The radial airflow through the small peaks and valleys of the peripheral air disc causes an unstable interface of differential viscosity to form finger-like structures (Fig. 6(b)). These finger-like structures, caused due to Saffman Taylor instability, propagate and branches into a fractal-like structure that we observe from the experimental data. Fig. 6(c) shows a schematic bottom cross-sectional view of air forming finger-like fractal structures in the P-P plane. 
The radial air layer thickness profile $h(r)$ obtained from the interference data is plotted for different pool viscosities for the same impact Weber number ($We{\sim}10$) in Fig. 7(a).
Drop impact on $5cSt$ has the highest air layer thickness followed by $350cSt$ and $950cSt$. The thickness profile shows detectable perturbations for $350cSt$ and $950cSt$ in the peripheral air disc region ($225{\mu}m<r<450{\mu}m$); however, negligible perturbations were observed for $5cSt$ (Fig. 7(b)). The perturbations observed in $350cSt$ were more prominent (larger in amplitude) than in $950cSt$. 
We note (Fig. 3 and Fig. 7(b)) that the peripheral air disc width is largest for $950cSt$, followed by $350cSt$ and $5cSt$, respectively.
Further, we also observe that the time scale for air disc to rupture is largest for $950cSt$ followed by $350cSt$ and $5cSt$ (Fig. 3). The air layer for $950cSt$ is comparatively more stable compared to $350$ and $5cSt$.
The characteristic length scales (Fig. 7(a) and 7(b)) relevant to the dynamics is the air layer thickness in the dimple region ($h_0{\sim}4-6{\mu}m$), and the peripheral air disc region ($h{\sim}0.3-0.4{\mu}m$). The distinct air layer dynamics and dewetting characteristics observed are based on the implications of thin-film/spinodal and Saffman-Taylor stability characteristics in the peripheral disc region discussed in the following sections.
\begin{figure*}
  \centerline{\includegraphics{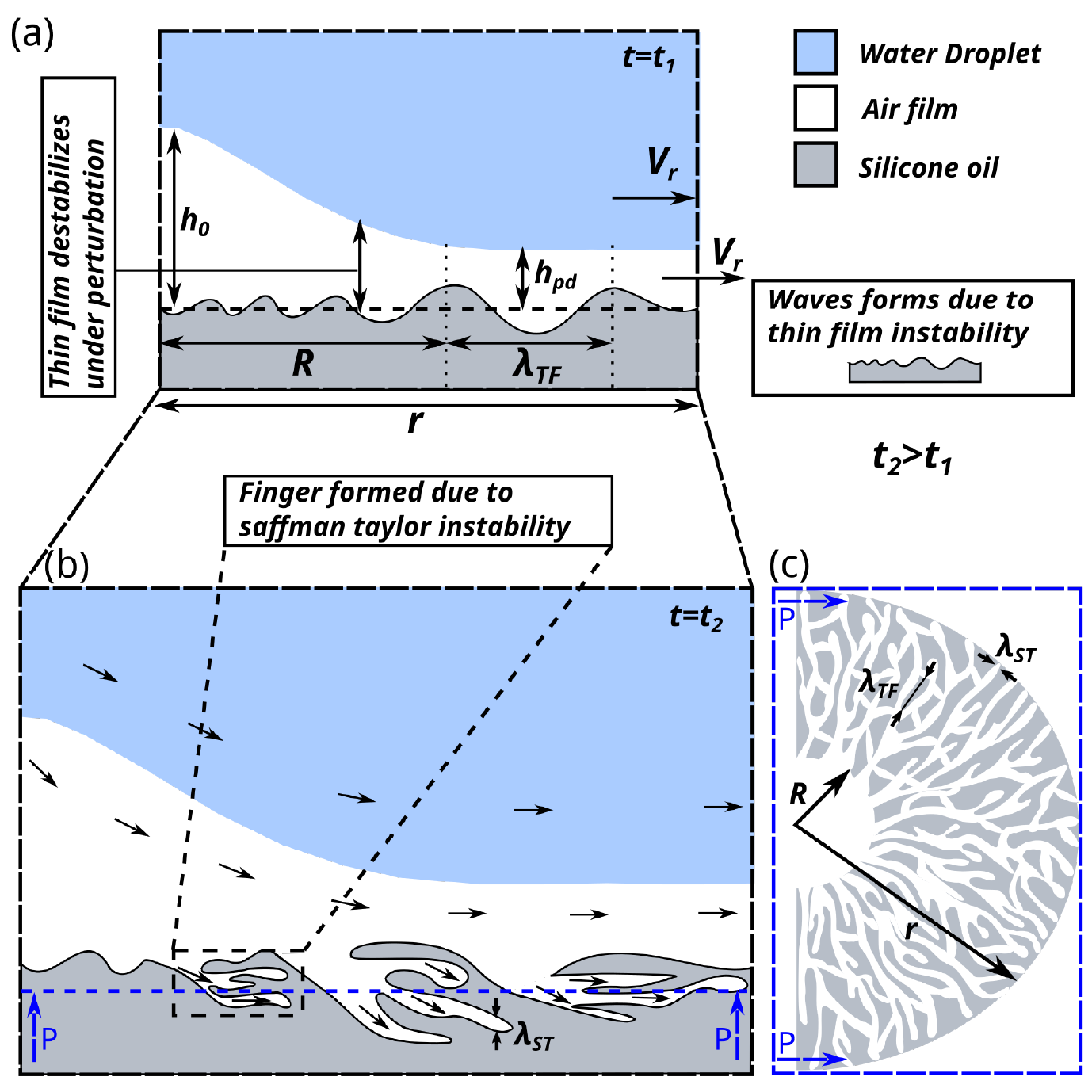}}
  \caption{Schematic representation depicting the mechanism of structure formation in peripheral air disc region. (a) Schematic depicting thin film instability and associated length scale fluctuations. (b) Schematic depicting the mechanism of Saffman Taylor instability leading to viscous fingering. (c) PP Cross sectional view of radial air expansion causing finger like structures to form and propagate as a fractal like structures in the radial direction.}
\label{Figure6}
\end{figure*}

\subsubsection{Thin film/Spinodal instability}
The squeezing of the entrapped air continues as the droplet sinks in the pool. As the thickness becomes smaller than a micron ($h<1{\mu}m$), Van der Walls force and double-layer interaction become comparable to capillary force, and hence new length scales can exist in metastable state due to the balance of intermolecular and capillary forces [18,38] before rupture initiates. Following the reasoning given by Vrij [18], the most dominant characteristic length scale due to intermolecular and capillary interaction for spinodal instability is given by ${\lambda}_{TF}$
\begin{equation}
    {\lambda}_{TF}{\sim}h^2{\left(\frac{{\sigma}_{aw}}{A}\right)^{1/2}}
\end{equation}
where $h$ is the air layer thickness in the peripheral air disc, $A$ is the Hamaker constant.
Using $h=h_{pd}$ from equation (4.18) in (4.19) we have
\begin{equation}
    {\lambda}_{TF}{\sim}{\left[\frac{AR_0}{{\sigma}_{aw}}\right]^{2/3}{\left[\frac{{\sigma}_{aw}}{A}\right]^{1/2}}}={\left[\frac{AR_0^4}{{\sigma}_{aw}}\right]^{1/6}}
\end{equation}

The experimental probability density function (PDF) of the length scales in the peripheral air disc region (given by the arithmetic mean of $a,b$ (Fig. 4(d)) of 63 measurements) is plotted in Fig. 7(c) (marked as Obs.) using kernel density estimate [39]. The experimental peak is approximately at $11{\mu}m$ which coincides remarkably to the scale predicted by equation (4.20).
${\lambda}_{TF}$ calculated through equation (4.20) using $A{\sim}\mathcal{O}(10^{-19}J)$, $R_0{\sim}\mathcal{O}(1.1{\times}10^{-3}mm)$  gives ${\lambda}_{TF}{\sim}11.26{\mu}m$.
Below a certain critical thickness, ${\lambda}_{TF}{\sim}h^2$ which indicates that small value of thickness corresponds to small ${{\lambda}_{TF}}$. This trend is consistent with our observations for $350cSt$ and $950cSt$. The spread of the PDF in fig. 7(c) is due to the multimodal nature of thin film instability generated due to height fluctuations. 

\begin{figure}
  \centerline{\includegraphics{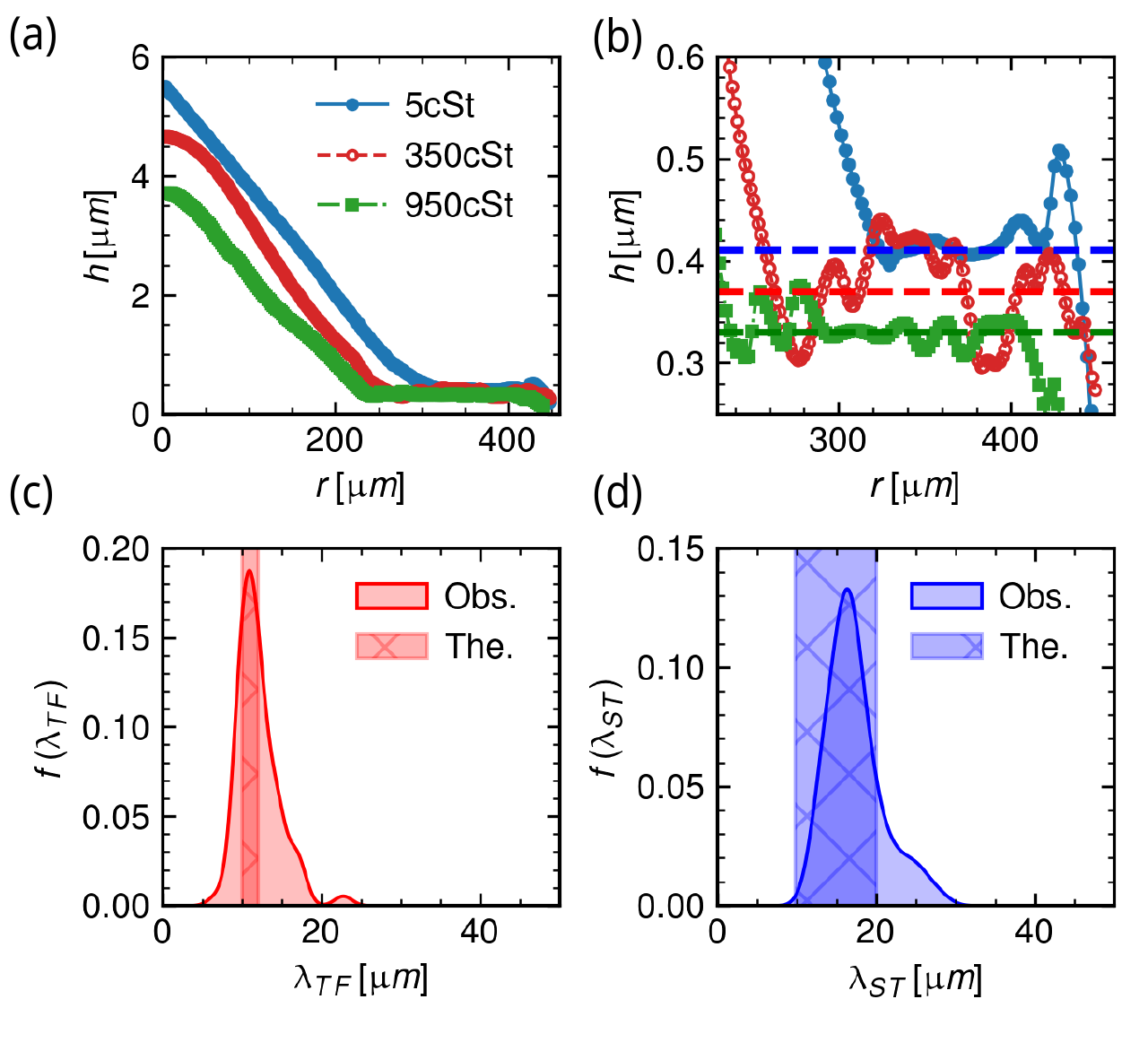}}
  \caption{(a) Air layer thickness profile $h(r)$ plotted as a function of radial distance for three different liquid pool viscosities ($5$, $350$ and $950cSt$). 
    (b) Air layer thickness profile in the peripheral air disc region.
    (c) The probability density function of the experimental length scales observed in the peripheral air disc region for $350cSt$ compared with the theoretical wavelength estimated from thin-film instability analysis.
    (d) Probability density function of the finger width $c$ observed in the peripheral air disc region for $350cSt$ compared with the theoretical estimates of the most dominant wavelength from Saffman Taylor instability theory.}
\label{Figure7}
\end{figure}

\subsubsection{Saffman Taylor Instability}
Fig. 6(b) and 6(c) depicts the mechanism of Saffman Taylor instability that causes finger like structures to form and propagate in the radial direction across an interface of viscosity gradient. The differential viscosity interface becomes unstable when a less viscous fluid (air here) displaces a high viscous fluid (Silicone oil). The most dominant unstable wavelength due to Saffamn Taylor instability in a radial Hele Shaw configuration is given by [40]
\begin{equation}
    {\lambda}_{ST}={2\sqrt{3}{\pi}r}{\left(\frac{Qr}{2{\pi}M{\sigma}_{as}}+1\right)^{-1/2}}
\end{equation}
where $r$ is a radial coordinate, $Q$ is the volume flow rate of air per unit depth, $M$ is the mobility of the air through the viscous silicone oil pool and ${\sigma}_{as}$ is the air-silicone oil surface tension. The volume flow rate through the peripheral air disc becomes
\begin{equation}
    Qh_{pd}=V2{\pi}rh_{pd}
\end{equation}
On simplification equation (4.22) becomes
\begin{equation}
    Q=V2{\pi}r
\end{equation}
where $V$ is the velocity at the interface of air silicone oil forming the fingers.
The mobility of air through silicone oil is given as [40]
\begin{equation}
    M=\frac{h_{pd}^2}{12{\mu}_s}
\end{equation}
Incorporating equation (4.23) and (4.24) in (4.21), the most dominant Saffman-Taylor wavelength scale can be written as
\begin{equation}
    {\lambda}_{ST}={2\sqrt{3}{\pi}r}{\left(\frac{12{\mu}_{s}r^2V}{{\sigma}_{as}h_{pd}^2}+1\right)^{-1/2}}
\end{equation}
where the $V$ is the radial velocity of the air layer across the silicone oil interface. The radial air velocity scale $V$ is given as [40]
\begin{equation}
    V=M\bm{\nabla}p{\sim}\frac{h_{pd}^2{\sigma}_{aw}}{12{\mu}_aR_0r}
\end{equation}
Substituting the scale of $V$ from equation (4.26) in equation (4.25) we have
\begin{equation}
    {\lambda}_{ST}{\sim}{2\sqrt{3}{\pi}r}{\left(\frac{{\mu}_s}{{\mu}_a}\frac{{\sigma}_{aw}}{{\sigma}_{as}}\frac{r}{R_0}+1\right)^{-1/2}}
\end{equation}
The experimental probability density function (PDF) of finger width $c$ calculated using kernel density estimate [39] for 67 measurements is shown in Fig. 7(d) (marked as Obs.).
The peak of the experimental PDF ($c{\sim}15{\mu}m$) lies within the theoretical limits (marked as The.) (${\lambda}_{ST}{\sim}9-20{\mu}m$) predicted by Saffman-Taylor instability theory [25].
The theoretical scaling for the finger width was calculated using equation (4.27) using a range of $50{\mu}m<r<200{\mu}m$, and it conforms with the experimental range. Fig. 7(d) shows that the experimental PDF distribution has a range from $8{\mu}m$ to $30{\mu}m$. The distribution is multimodal in nature due to the dependence of ${\lambda}_{ST}$ on the radial coordinate $r$.

Non-dimensionalizing Equation (4.27)  based on the dimple radius $R$ we have
\begin{equation}
    \frac{{\lambda}_{ST}}{R}{\sim}2\sqrt{3}{\pi}{\left(\frac{{\mu}_s}{{\mu}_a}\frac{{\sigma}_{aw}}{{\sigma}_{as}}\frac{R}{R_0}+1\right)^{-1/2}}
\end{equation}

\begin{figure}
  \centerline{\includegraphics{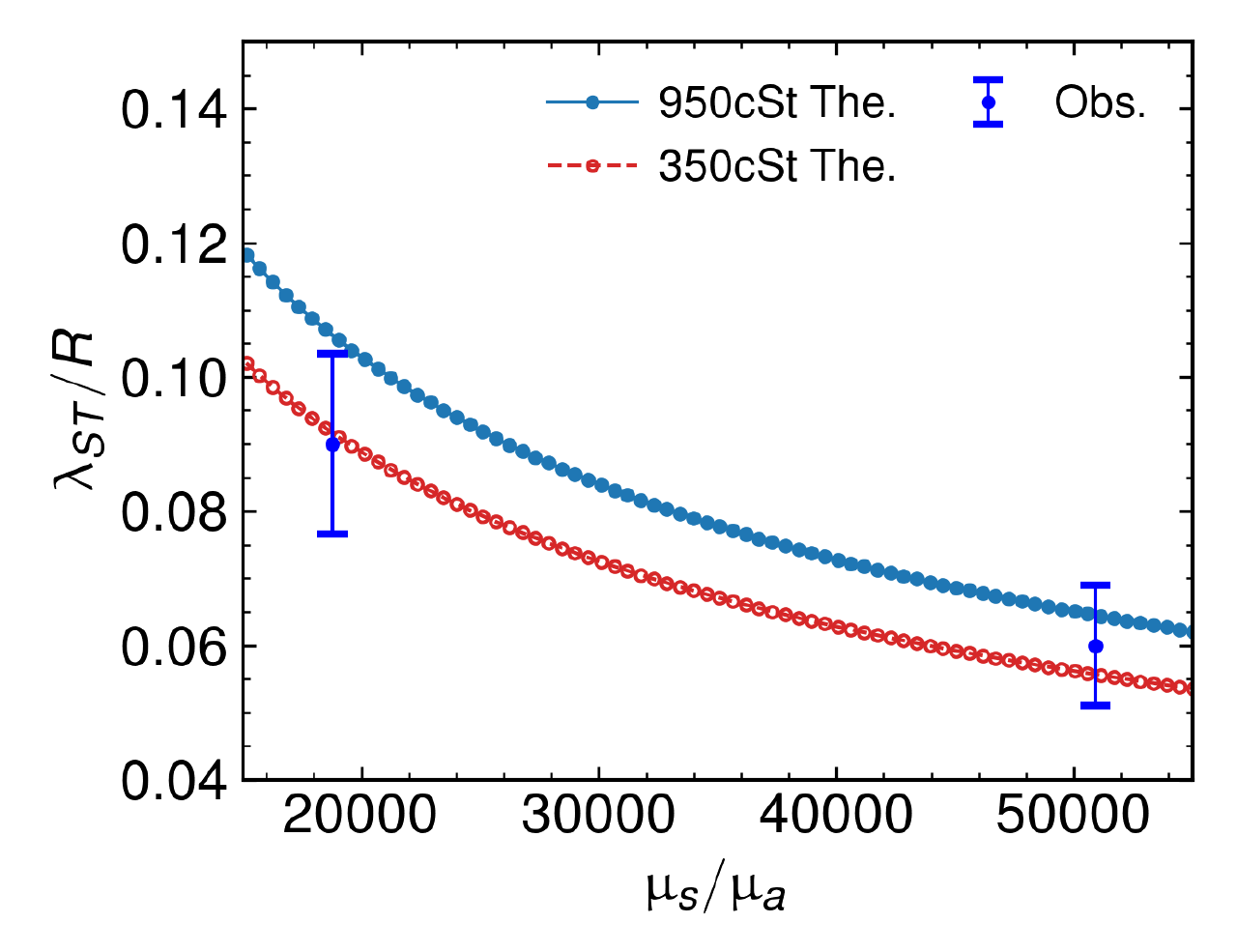}}
  \caption{Experimental and theoretical comparison depicting the dependence of Saffman Taylor wavelength (${\lambda}_{ST}$ as a function of viscosity ratio (${\mu}_s/{\mu}_a$)) for $350$ and $950cSt$.}
\label{Figure8}
\end{figure}

Theoretically, ${\lambda}_{ST}$ depends on the viscosity ratio, surface tension ratio, and the radial coordinate (equation (4.27 and 4.28)). Fig. 8 depicts the dependence of dimensionless Saffman Taylor length scale ${\lambda}_{ST}/R$ on the viscosity ratio (${\mu}_s/{\mu}_a$).
The characteristic length scale has an inverse dependence on the viscosity ratio, indicating that as the viscosity ratio reduces, the length scale increases, which is consistent for our cases (Fig. 3 and Fig. 7). The experimental observation conforms with the theoretical scaling predicted by equation (4.27) on an order of magnitude sense.
It is to be noted that the azimuthal and the radial length scales are approximately of the same order of magnitude (Fig. 7(c) and 7(d)). This is due to the fact that, thin film instability causes the air-silicone oil interface to develop perturbations, which further acts as a scaffolding for Saffman-Taylor instability to form finger like structures that grows and propagates as fractal like structure in the radial direction.



\section{Conclusion}
In conclusion, we observed that the central air dimple thickness $h_0$ and the peripheral air disc expansion was fastest for $5cSt$ followed by $350$ and $950cSt$. On the contrary, the air disc for $950cSt$ is the most stable, followed by $350$ and $5cSt$ indicating largest rupture time scale for $950cSt$. Further, it was also shown that the peripheral air disc radius $r_0$ grows as $t^{2/3}$ due to very small capillary number ($Ca{\sim}\mathcal{O}(10^{-4})$).
Unique hydrodynamic structures in the peripheral air disc during drop impact on immiscible liquid pool of $350$ and $950cSt$ at low impact Weber number ($We{\sim}10$) was observed. Morphology of the structures shows that the characteristic wavelength observed aligns with the predictions of thin-film/spinodal (${\lambda}_{TF}{\sim}\mathcal{O}(11.26{\mu}m)$) and Saffman Taylor instability ${\lambda}_{ST}{\sim}\mathcal{O}(13.89{\mu}m)$ length scales.

\vskip6pt

\enlargethispage{20pt}


\dataccess{A zip file name \href{https://drive.google.com/file/d/1p_a8bppg7aVLDHQDb3JuG-JhQQoOXPDZ/view?usp=sharing}{SD1} is provided as a link carrying the data used in analysis. The link is embedded in the supplementary material.}

\aucontribute{DR and SM proposed the problem statement. DR, SM, SRS performed the experiments and data analysis. DR and SB carried out the theoretical analysis and wrote the paper. SB supervised the project.}

\competing{The authors declare no competing interest.}

\funding{The authors are thankful for the funding received from the Defence Research and Development Organization  (DRDO) Chair Professorship.}




\end{document}